\pdfoutput=1
\documentclass[twocolumn,10pt]{article}
\usepackage[pdftex]{graphicx} 
\usepackage[font=footnotesize]{caption}
\usepackage[font=scriptsize]{subcaption}
\usepackage{setspace}
\usepackage{url}
\usepackage{multirow}
\usepackage{pifont} 
\usepackage{lipsum} 
\usepackage[numbers,sort&compress,sectionbib]{natbib} 
\usepackage{color}
\usepackage{floatrow}
\usepackage{subfig}
\usepackage{multirow}
\usepackage{array}
\usepackage{enumitem}
\usepackage{verbatim}

\newcommand{\dn}{{DroidNative}}

\begin{document}\sloppy
%

\title{\textbf{\dn: Semantic-Based Detection of Android Native Code Malware}}


\author{\hspace*{-10pt}
\begin{minipage}[t]{2.3in} \normalsize \baselineskip 12.5pt
\centerline{\textbf{Shahid Alam}}
\centerline{Qatar University}
\centerline{salam@qu.edu.qa}
\end{minipage} \kern 0in
\begin{minipage}[t]{2.3in} \normalsize \baselineskip 12.5pt
\centerline{\textbf{Zhengyang Qu}}
\centerline{Northwestern University}
\centerline{zhengyangqu2017@u.northwestern.edu}
\end{minipage}
\begin{minipage}[t]{2.3in} \normalsize \baselineskip 12.5pt
\centerline{\textbf{Ryan Riley}}
\centerline{Qatar University}
\centerline{rriley@qu.edu.qa}
\end{minipage}\\
\begin{minipage}[t]{2.3in} \normalsize \baselineskip 12.5pt
\centerline{\textbf{Yan Chen}}
\centerline{Northwestern University}
\centerline{ychen@northwestern.edu}
\end{minipage}
\begin{minipage}[t]{2.3in} \normalsize \baselineskip 12.5pt
\centerline{\textbf{Vaibhav Rastogi}}
\centerline{University of Wisconsin-Madison}
\centerline{vrastogi@wisc.edu}
\end{minipage}
}

\maketitle

\begin{abstract}

According to the Symantec and F-Secure threat reports, mobile malware development in 2013 and 2014 has continued to focus almost exclusively ({\raise.17ex\hbox{$\scriptstyle\sim$}}99\%) on the Android platform. Malware writers are applying stealthy mutations (obfuscations) to create malware variants, thwarting detection by signature based detectors. In addition, the plethora of more sophisticated detectors making use of static analysis techniques to detect such variants operate only at the bytecode level, meaning that malware embedded in native code goes undetected. A recent study shows that 86\% of the most popular Android applications contain native code, making this a plausible threat. This paper proposes \emph{{\dn}}, an Android malware detector that uses specific control flow patterns to reduce the effect of obfuscations, provides automation and platform independence, and as far as we know is the first system that operates at the Android \emph{native code} level, allowing it to detect malware embedded in both native code and bytecode. When tested with traditional malware variants it achieves a detection rate (DR) of 99.48\%, compared to academic and commercial tools' DRs that range from 8.33\% -- 93.22\%. When tested with a dataset of 2240 samples \emph{{\dn}} achieves a DR of 99.16\%, a false positive rate of 1.3\% and an average detection time of 26.87 sec/sample.

\end{abstract}

\textbf{Keywords} End point security, Android native code, Malware Analysis, Malware variants detection, Control flow analysis, Data mining

\section{Introduction}\label{sec:introduction}

Worldwide losses due to malware attacks and phishing between July 2011 and July 2012 were \$110 billion \cite{Symantec-2012}. In 2013 there was a 42\% increase in the malware attacks over 2011. Web-based attacks increased by 30 percent in 2012 \cite{Symantec-2012}. Mobile malware development in 2013 \cite{Symantec-2014} and in 2014 \cite{F-Secure-2014} continues to focus almost exclusively ({\raise.17ex\hbox{$\scriptstyle\sim$}}99\%) on the Android platform, due to its popularity and open-nature.

In recent years there has been a significant amount of research focusing on detection of Android malware through both static \cite{DroidSIFT-AM,DroidLegacy-AM,AndroSimilar,CBCFG-AM,Droidminer-AM,Dendroid-AM,DroidAnalyzer-AM,AsDroid-AM,DynamicCodeLoading-AM} and dynamic \cite{DroidRanger-AM,DynamicCodeLoading-AM,DroidAnalytics-AM,TaintDroid-AM,CrowDroid,DroidScope} analysis techniques. Among the static analysis techniques, all the systems operate at the Java bytecode level.  This is logical, as most Android applications are primarily written in Java. Many Android applications, however, contain functionality not implemented in Java but instead in \emph{native code}.

In this paper, we refer to native code as machine code that is directly executed by a processor; as opposed to bytecode, which is executed in a virtual machine such as the JVM or Dalvik-VM. While the standard programming model for Android applications is to write the application in Java and distribute the bytecode, the model also allows for developers to include native code binaries as part of their applications.  This is frequently done in order to make use of a legacy library or to write code that is hardware dependent, such as for a game graphics engine or video player.

A recent study~\cite{NativeGuard} shows that 86\% of the most popular Android applications contain native code. Given that existing, static analysis based malware detection systems operate on a bytecode level, this means these applications cannot be completely analyzed.  If the malicious payload of the app was included in the native code, existing systems would be unable to detect it.

In addition to native code, another issue plaguing malware detection on Android is obfuscation. To protect an Android application for reverse engineering attacks, even legitimate developers are encouraged to obfuscate~\cite{Obfuscation-Def-1} their code. Similar techniques are used by an Android malware writer to prevent analysis and detection. Obfuscation can be used to make the code more difficult to analyze, or to create \emph{variants} of the same malware in order to evade detection. Previous studies \cite{DroidChameleon,ADAM,Obfuscation-Faruki} have evaluated the resilience of commercial anti-malware products when tested against variants of known malware.  In~\cite{DroidChameleon}, the authors found that none of the products were able to detect the variants. In more recent work~\cite{ADAM} the detection rate of the products ranged from 50.95\% -- 76.67\%.

In order to try and detect malware variants, several works have focused on Android malware detection using static analysis. Some of these, such as~\cite{DroidSIFT-AM} and~\cite{DroidLegacy-AM}, use API call graphs (ACGs); \cite{CBCFG-AM} uses component-based ACGs; \cite{AndroSimilar} uses byte sequence features. Other previous works have focused on graphs to represent program semantics, such as control-flow graphs~\cite{Semantic-MD-2005} and data dependency graphs \cite{Malware-Specifications-2010,Efficient-MD-2009}. These techniques make an exact match against manually-crafted specifications for malware detection, and therefore can potentially be evaded by malware variants. Other semantic-based techniques are compute intensive, on reducing the processing time~\cite{Code-Graph-MD} they produce poor detection rates (DRs) and are not suitable for real-time malware detection.

This paper introduces {\dn}, a malware detection system for Android that operates at the \emph{native code level} and is able to detect malware in either bytecode or native code.  {\dn} performs static analysis of the native code and focuses on patterns in the control flow that are not significantly impacted by obfuscations. This allows {\dn} to effectively detect Android malware, even with obfuscations applied and even if the malicious payload is embedded in native code. {\dn} is not limited to only analyzing native code, it is also able to analyze bytecode by making use of the Android runtime (ART)~\cite{ART} to compile bytecode into native code suitable for analysis. The use of control flow with patterns enables {\dn} to detect smaller size malware, which allows {\dn} to reduce the size of a signature for optimizing the detection time without reducing the DR. Our experimental analysis shows that {\dn} is able to achieve an overall DR of 99.16\% and a false positive rate (FPR) of 1.3\%. {\dn} achieves an average detection time of 26.87 sec/sample when tested with 2240 samples, and is {\raise.17ex\hbox{$\scriptstyle\sim$}}6.5 times faster than the other compared work~\cite{DroidSIFT-AM}. \\
The contributions of this work are as follows:

\begin{itemize}[leftmargin=*]
	\item {\dn}, as far as we know, is the first system that builds and designs cross-platform signatures for Android and operates at the native code level, allowing it to detect malware embedded in either bytecode or native code.
	\item When applied to traditional, bytecode based Android malware, the detection rate is on par with that of many research systems that only operate at the bytecode level.
	\item When tested with previously unseen variants of existing malware, the detection rate is 99.48\%, significantly surpassing that of commercial tools.
	\item Based on our analysis of related work, {\dn} is noticeably faster than existing systems, making it suitable for real-time analysis.
	\item We provide n-fold cross validation results for the system, providing a more systematic approach to demonstrating accuracy than is employed by most similar works.
\end{itemize}

\section{Background}
\label{sec:binary-analysis}

{\dn} detects malware by building control flow (behavioral) signatures of an
application and comparing them to signatures derived from known malware.
In order to build these signatures, the Android program executable (provided in either bytecode or native code) is converted to an intermediate representation. 
This intermediate representation provides automation, platform independence and reduces the effect of obfuscations. 

In this section we provide background information regarding the intermediate representation used as well as the Android Runtime Environment (ART) that are
used in the design and implementation of {\dn}.

\subsection{Malware Analysis Intermediate Language (MAIL)}\label{sec:mail}\label{sec:acfg-swod}

{\dn} performs its analysis on an intermediate representation of an Android
application's binary code. In order to accomplish this, {\dn} makes use of the Malware Analysis Intermediate Language (MAIL)~\cite{MAIL-2013}. Traditionally, intermediate languages are used in compilers to translate the source code into a form that is easy to optimize and to provide portability. We can apply these same concepts to malware analysis and detection. MAIL provides a high level representation of the disassembled binary program including specific information such as control flow information, function/API calls and patterns etc.  Having this information available in an IL allows for easier and optimized analysis and detection of malware.

{\dn} uses MAIL to provide an abstract representation of an assembly program, and that representation is used for malware analysis and detection. By translating native binaries compiled for different platforms (Android supports ARM, x86 and MIPS architectures) to MAIL, {\dn} achieves platform independence, and uses MAIL Patterns to optimize malware analysis and detection.

\subsubsection{Patterns for Annotation}\label{sec:mail-patterns}
MAIL is used to annotate a CFG (control flow graph) of a program using different patterns available in the language. The purpose of these annotations is to assign patterns to MAIL statements that can be used later for pattern matching during malware detection, that helps us detect smaller malware, and enables us to reduce the size of a signature and the detection time.

There are a total of eight basic statements (e.g, assignment, control and conditional, etc) in MAIL that can be used to represent the structural and behavioral information of an assembly program. There are 21 Patterns in the MAIL language, and every MAIL statement can be tagged by one of these patterns. A pattern represents the type of a MAIL statement and can be used for easy comparison and matching of MAIL programs. For example, an assignment statement with a constant value and an assignment statement without a constant value are two different patterns. The details of the patterns can be found in Table~\ref{tab:mailpatterns}.

\begin{table*}[!tb]
\centering
\caption{Patterns used in MAIL.  $r_0$, $r_1$, $r_5$ are the general purpose registers, $zf$ and $cf$ are the zero and carry flags respectively, and $sp$ is the stack pointer.}
{
\small
\begin{tabular}{|l|p{13.2cm}|}
\hline
\textbf{Pattern} & \textbf{Description} \\
\hline
$\mathtt{ASSIGN}$ & An assignment statement, \emph{e.g.} $r_0$ = $r_0$ + $r_1$; \\
$\mathtt{ASSIGN\_CONSTANT}$ & An assignment statement including a constant, \emph{e.g.} $r_0$ = $r_0$ + 0x01; \\
$\mathtt{CONTROL}$ & A control statement where the target of the jump is unknown, \emph{e.g.} if (zf == 1) JMP [$r_0$ + $r_1$ + 0x10]; \\
$\mathtt{CONTROL\_CONSTANT}$ & A control statement where the target of the jump is known. \emph{e.g.} if ($zf$ == 1) JMP 0x400567; \\
$\mathtt{CALL}$ & A call statement where the target of the call is unknown, \emph{e.g.} CALL $r_5$; \\
$\mathtt{CALL\_CONSTANT}$ & A call statement where the target of the call is known, \emph{e.g.} CALL 0x603248; \\
$\mathtt{FLAG}$ & A statement including a flag, \emph{e.g.} $cf$ = 1; \\
$\mathtt{FLAG\_STACK}$ & A statement including flag register with stack, \emph{e.g.} $eflags$ = [$sp$ = $sp$ -- 0x1]; \\
$\mathtt{HALT}$ & A halt statement, \emph{e.g.} HALT; \\
$\mathtt{JUMP}$ & A jump statement where the target of the jump is unknown, \emph{e.g.} JMP [$r_0$ + $r_5$ + 0x10]; \\
$\mathtt{JUMP\_CONSTANT}$ & A jump statement where the target of the jump is known, \emph{e.g.} JMP 0x680376 \\
$\mathtt{JUMP\_STACK}$ & A return jump, \emph{e.g.} JMP [$sp$ = $sp$ -- 0x8] \\
$\mathtt{LIBCALL}$ & A library call, \emph{e.g.} compare($r_0$,$r_5$); \\
$\mathtt{LIBCALL\_CONSTANT}$ & A library call including a constant, \emph{e.g.} compare($r_0$,0x10); \\
$\mathtt{LOCK}$ & A lock statement, \emph{e.g.} lock; \\
$\mathtt{STACK}$ & A stack statement, \emph{e.g.} $r_0$ = [$sp$ = $sp$ -- 0x1]; \\
$\mathtt{STACK\_CONSTANT}$ & A stack statement including a constant, \emph{e.g.} [$sp$ = $sp$ + 0x1] = 0x432516; \\
$\mathtt{TEST}$ & A test statement, \emph{e.g.} $r_0$ and $r_5$; \\
$\mathtt{TEST\_CONSTANT}$ & A test statement including a constant, \emph{e.g.} $r_0$ and 0x10; \\
$\mathtt{UNKNOWN}$ & Any unknown assembly instruction that cannot be translated. \\
$\mathtt{NOTDEFINED}$ & The default pattern, \emph{e.g.} all the new statements when created are assigned this default value. \\
\hline
\end{tabular}
}
\label{tab:mailpatterns}
\end{table*}

\subsubsection{Signature Generation Techniques}\label{sec:sgt}
To counter simple signature-based malware detectors, malware will often use
obfuscation techniques~\cite{DroidChameleon,ADAM,Obfuscation-Faruki} to obscure its host application and make it difficult to understand, analyze and detect the malware embedded in the code. In order to detect malware protected
by obfuscations, we apply two signature generation techniques designed
to counter such approaches. MAIL Patterns are used to build these signatures.

\paragraph{Annotated Control Flow Graph}

The first technique, Annotated Control Flow Graph (ACFG), analyzes a control flow graph derived from an application and
then annotates it using MAIL patterns in order to capture the control flow semantics of the program. A complete formal definition of an ACFG and its matching for malware detection is given in~\cite{ACFG-2014}. 

Annotating the CFG with MAIL statements allows much smaller CFGs to be reliably analyzed and used for malware detection than are traditionally used in CFG based techniques.  The advantage of this is that an application can be broken up into a collection of smaller ACFGs instead of one, large CFG. This collection can then be analyzed and pattern matched with malware much more quickly than the larger CFG.

For example, an application being analyzed is broken up into a set of small ACFGs which become that application's signature.  At detection time, the ACFGs within that signature are compared to the ACFGs of known malware, and if a high percentage of ACFGs in the application match that of known malware, then the application can be flagged as malware.  This matching is also very fast, because the individual ACFGs being compared against are fairly small.

\paragraph{Sliding Window of Difference}

The second technique for signature building, Sliding Window of Difference (SWOD)~\cite{SWOD-CFWeight-2014}, is an opcode-based technique for malware detection. One of the main advantages of using opcodes for detecting malware is that detection can be performed in real-time. However, the current techniques \cite{Opcode-Graph-MD,Opcode-Histogram,Opcode-HMM-MD,Opcode-SD-MD} using opcodes for malware detection have several disadvantages: 
(1) The patterns of opcodes can be changed by using a different compiler, the same compiler with a different level of optimizations, or if the code is compiled for a different platform. 
(2) Obfuscations introduced by malware variants can change the opcode distributions. 
(3) The execution time depends on the number of features selected for mining in a program. Selecting too many features results in a high detection rate but also increases the execution time. Selecting too few features has the opposite effect.

A SWOD is a window that represents differences in MAIL patterns' (while provide a high-level representation of a program's opcodes) distributions and hence makes the analysis independent of different compilers, compiler optimizations, instruction set architectures and operating systems. In order to mitigate the effect of obfuscations introduced by malware variants, SWOD uses a control flow weight scheme based on heuristics~\cite{SWOD-CFWeight-2014} to extract and analyze the control flow semantics of the program. Furthermore, it uses statistical analysis of MAIL patterns' distributions to develop a set of heuristics that help in selecting an appropriate number of features and reduce the runtime cost.

\subsection{The ART Runtime Environment}

Android applications are available in the form of APKs (Android application packages). An APK file contains the application code as Android bytecode and precompiled native binaries (if present). Traditionally, those applications are executed on the phone using a customized Java virtual machine called Dalvik~\cite{DALVIK}.  Dalvik employs just-in-time compilation (JIT), similar to the standard JVM.

Starting from Android 5.0, the Dalvik-VM was replaced by the ART~\cite{ART}. ART uses ahead of time (AOT) compilation to transform the Android bytecode into native binaries
when the application is first installed on the device. ART's AOT compiler can perform complex and advanced code optimizations that are not possible in a virtual machine using a JIT compiler. ART improves the overall execution efficiency and reduces the power consumption of an Android application, but increases the installation time and may increase the code size of an application.
In this work, we use ART in order to translate bytecode into native code
that can be analyzed using our techniques.

\section{System Design}\label{sec:design}

\begin{figure*}[tbp]
	\centering
	\includegraphics[width=0.75\textwidth]{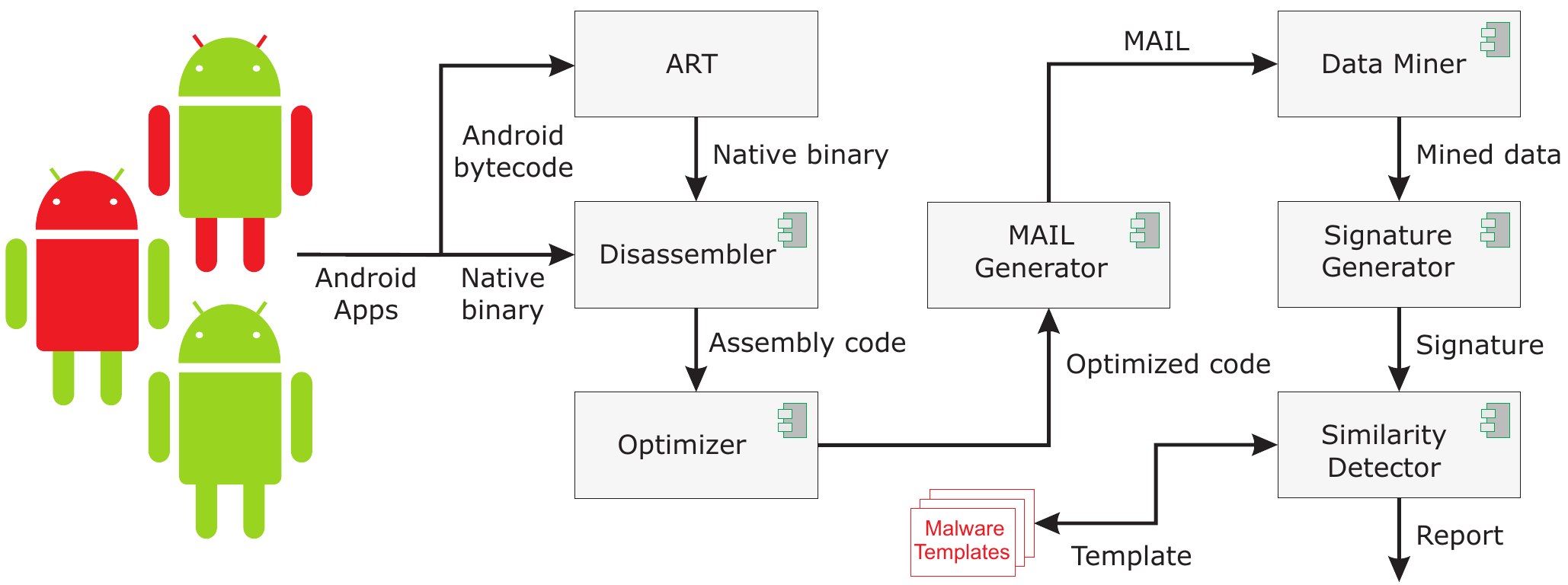}
	\caption{Overview of {\dn}.}
	\label{fig:DroidNative}
\end{figure*}

In this section we present the design of {\dn}. {\dn} provides a fully automated approach to malware detection based on analysis of the native code of an Android application. It analyzes control-flow patterns found in the code in order to compare them to patterns seen in known malware.

Figure~\ref{fig:DroidNative} provides an overview of the {\dn} architecture. As mentioned previously, the system operates by analyzing native code.  The vast majority of Android applications, however, are distributed as bytecode. (For example, a standard application might include the main application as Java bytecode, but also include native code libraries for performance reasons.) When an app is presented to {\dn} for analysis, the application is separated into bytecode and native code components. Any bytecode components are passed to the ART compiler and a native binary is produced. Any native code components are passed directly to the next stage. At this point, all parts of the application exist as native code.

Next, the binary code is disassembled and translated into MAIL code, which is processed using ACFG and SWOD (more details can be found below) in order to produce a \emph{behavioral signature}.  Next, the \emph{similarity detector} uses this signature and detects the presence of malware in the application using the \emph{malware templates} previously discovered during a training phase. A simple binary classifier (decision tree) is used for this purpose. If the application being analyzed matches a malware template (within a given threshold), then the app is tagged as malware.

We will now discuss the various components of {\dn} in more detail.

\subsection{Disassembler}
When provided with the native code for an application, the first step
in the {\dn} process is to disassemble it.
A challenge related to this stage is ensuring that all code is found and
disassembled.  
There are two standard techniques \cite{Disassembly-Binary-2002} for disassembly. (1) The \emph{Linear Sweep} technique starts from the first byte of the code and disassembles one instruction at a time until the end. This assumes that the instructions are stored adjacently, and hence does not distinguish between embedded data and instructions. When data is mixed with the code either by the compiler or by malware writers, this technique may produce incorrect results. The advantage of this technique is that it provides complete coverage of the code. (2) The \emph{Recursive Traversal} technique decodes instructions by following the control flow of the program. This technique only disassembles an instruction if it is referenced by another instruction. The advantage of this technique is that it distinguishes code from data. But in case of a branch whose destination address cannot be determined statically, this technique may fail to find and disassemble valid instructions. To overcome the deficiencies of linear sweep and recursive traversal we combine these two techniques while disassembling. For example, in case of an unknown branch, we use MAIL to tag such address, and use a linear sweep to move to the next instruction.

Another challenge related to this is that most binaries used in Android are stripped, meaning they do not include debugging or symbolic information.  This makes it extremely difficult to statically build a call (API) graph for such binaries. It can also become difficult to find function boundaries. IDA Pro \cite{IDA-Pro}, the most popular disassembler, uses pattern matching to find such functions, but these patterns change from one compiler to another and also from one version to another version of the same compiler. Therefore, this technique is difficult to maintain and is not general. We handle this problem by building control flow patterns and use them for malware detection. Our focus is mostly on the detection of malware variants, therefore, we may only need to find where the control is flowing (i.e: just the behavior and not the function boundaries), and then compare this behavior with the previous samples of malware available to detect such malware.


\subsection{Optimizer}

The Optimizer performs normalization of the assembly code. The normalizations performed are the removal of NOP, junk and some of the prefixes. The Optimizer also prepares the assembly code to be translated to MAIL by removing other instructions that are not required for malware analysis.


A key challenge for static analysis of native code is the 
large number of instructions supported by the Intel x86 and ARM microprocessors (the two most popular architectures supported by Android).
There are hundreds of different instructions in instruction sets for these architectures. In order to speedup static analysis, we need to reduce the number of these instructions that we fully process.
Many instructions are not relevant to malware analysis. Examples include the instruction PREFETCH in Intel x86 and PRFM in ARM. These instructions move data from the memory to the cache. 
The optimizer removes these and other similar instructions.

An Android application is divided into components, and each component serves a different purpose, and is called when an event occurs.  In addition, exceptions permeate the control flow structure of an Android app. Exceptions are resolved by dynamically locating the code specified by the programmer for handling the exception. This produces several unknown branch addresses in a disassembled Android binary program. The existence of multiple entry points makes it difficult to build a correct control flow graph~\cite{Dragon-Book} for the program using static analysis. To handle this problem, {\dn} builds multiple, smaller, interwoven CFGs for a program instead of a single, large CFG. These smaller CFGs are then used for matching and malware detection. Matching smaller CFGs, instead of a single large CFG, also helps in reducing the runtime.

\subsection{MAIL Generation}

The MAIL Generator translates an assembly program to a MAIL program. Some of the major tasks performed by this component are: (1) Translating each assembly instruction to the corresponding MAIL statement(s). Some of the assembly instructions, such as PUSHA (x86) and STMDB (ARM), are translated to more than one MAIL statement. (2) Assigning a pattern to each MAIL statement. (3) Reducing the number of different instructions by grouping together functionally equivalent assembly instructions so that the MAIL Generator can combine them into one MAIL statement using the MAIL Patterns listed in Table~\ref{tab:mailpatterns}.

Further details regarding the transformation of Intel x86 and ARM assembly programs into MAIL can be found in our technical report~\cite{TR-MAIL}.





\subsection{Malware Detection}

In this Section we explain how the components, \textbf{Data Miner}, \textbf{Signature Generator} and \textbf{Similarity Detector}, shown in Figure \ref{fig:DroidNative}, use the two techniques, ACFG and SWOD, for malware detection. The Data Miner searches for the control and structural information in a MAIL program to help the Signature Generator build a behavioral signature (currently two types of signatures, ACFG or SWOD) of the MAIL program. The Similarity Detector matches the signature of the program against the signatures of the malware templates extracted during the training phase, and determines
whether the application is malware based on thresholds that are computed empirically as explained in Section~\ref{sec:mt}, for malware detection.

\subsubsection{{\dn}-ACFG}\label{sec:acfg}

For detecting Android malware variants using ACFGs, a CFG is built for
each function in the annotated MAIL program, yielding the ACFGs. These ACFGs, i.e, the signature of the Android program, are matched against the malware templates to see if the program contains malware or not. If a high number of ACFGs within a program are flagged as malware, then the application itself is flagged.

\begin{figure}[tb]
\centering
\begin{subfigure}{0.25\textwidth}
   \centering
   \includegraphics[scale=0.25]{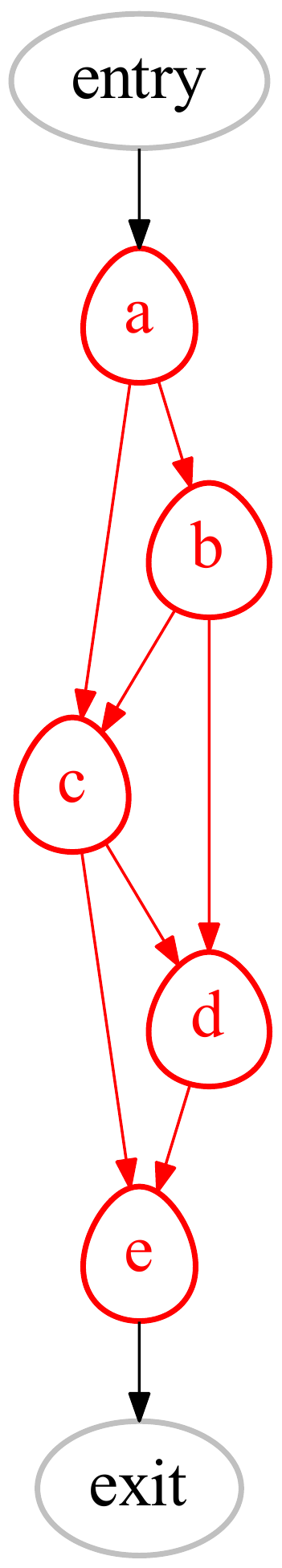}
\end{subfigure}
\begin{subfigure}{0.25\textwidth}
   \centering
   \includegraphics[scale=0.25]{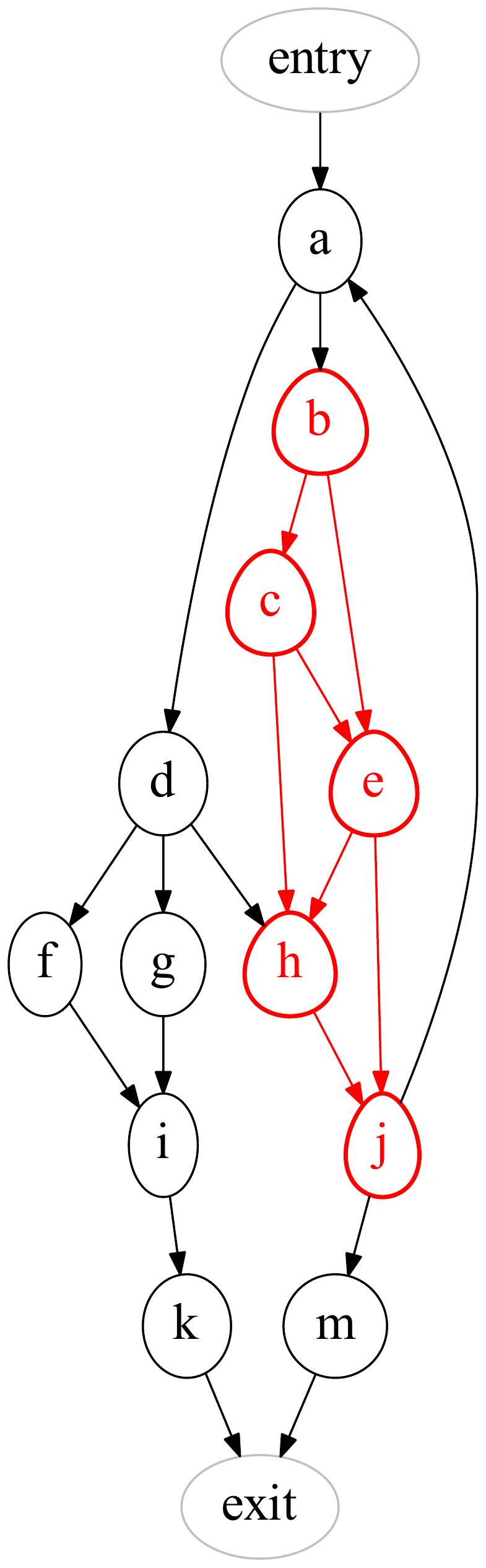}
\end{subfigure}
\caption{An example of ACFG matching. The graph on the left ($M$, a malware sample) is matched as a subgraph of the graph on the right ($P_{sg}$, the malware embedded inside a benign program), i.e, $M = (a,b,c,d,e) \cong P_{sg} = (b,c,e,h,j)$, where $M \cong P_{sg}$ denotes that $M$ is isomorphic \normalfont{\cite{Graph-Theory}} to $P_{sg}$.}
\label{fig:subgraph-matching}
\end{figure}

If a part of the control flow of a program contains a malware, we classify that program as malware, i.e, if a percentage (compared to some predefined threshold) of the number of ACFGs in the signature of a malware program match with the ACFGs in the signature of a program, then the program is classified as malware. An example of ACFG matching can be found in Fig.~\ref{fig:subgraph-matching}.

\begin{figure*}[tb]
\centering
{\includegraphics[scale=0.50]{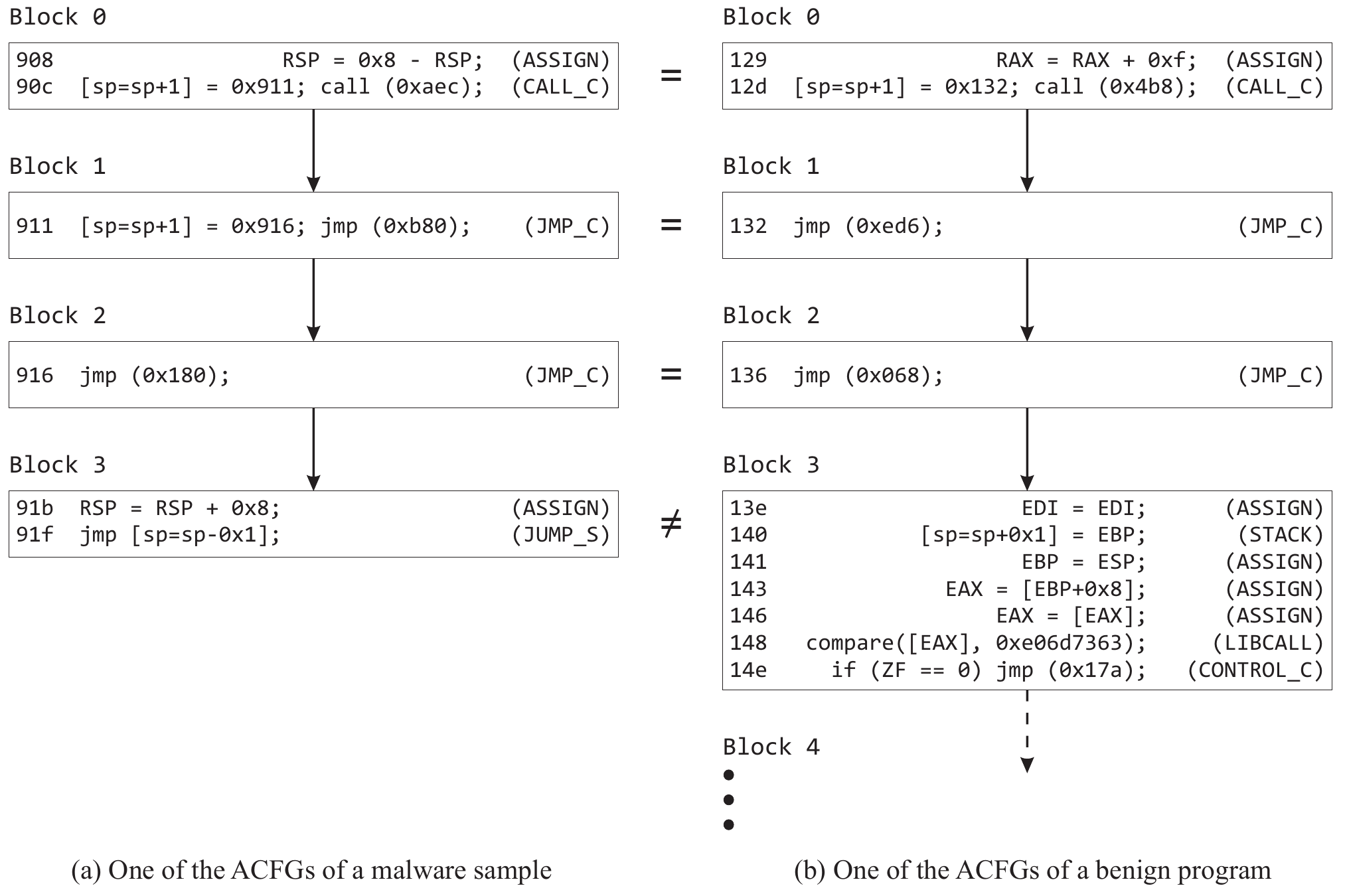}}
\caption{Example of a failed \emph{pattern matching} of two \emph{isomorphic} ACFGs. The ACFG on the left (a malware sample) is \emph{isomorphic} to the subgraph (blocks 0 -- 3) of the ACFG on the right (a benign sample).}
\label{fig:pattern-matching-example}
\end{figure*}

If an ACFG of a malware program matches with an ACFG of a program, we further use the patterns to match each statement in the matching nodes of the two ACFGs. A successful match requires all the statements in the matching nodes to have the same patterns, although there could be differences in the corresponding statement blocks. Example of a failed \emph{pattern matching} of two \emph{isomorphic} ACFGs is shown in Figure \ref{fig:pattern-matching-example}. This allows us to detect malware with smaller ACFGs, and also allows us to reduce the size (number of blocks) of an ACFG for optimizing the detection time. We reduce the number of blocks in an ACFG by merging them together. Two blocks are combined only if their merger does not change the control flow of the program.

\subsubsection{{\dn}-SWOD}\label{sec:swod}

Every MAIL statement is assigned a pattern during translation from assembly language to MAIL. Each MAIL pattern is assigned a weight based on the SWOD that represents the differences between malware and benign samples' MAIL patterns' distributions. 
The CFWeight (control flow weight) of a MAIL statement is computed by adding all the weights assigned to the elementary statements involved in it using some heuristics. The final weight of a MAIL statement is the sum of its CFWeight and the weight of the pattern assigned to the statement. The final weights of the statements of a MAIL program are stored in a weight vector that represents the program signature. This signature is then sorted in ascending order for easy and efficient comparison using an index-based array.

\begin{figure}[tb]
\centering
{\includegraphics[scale=0.40]{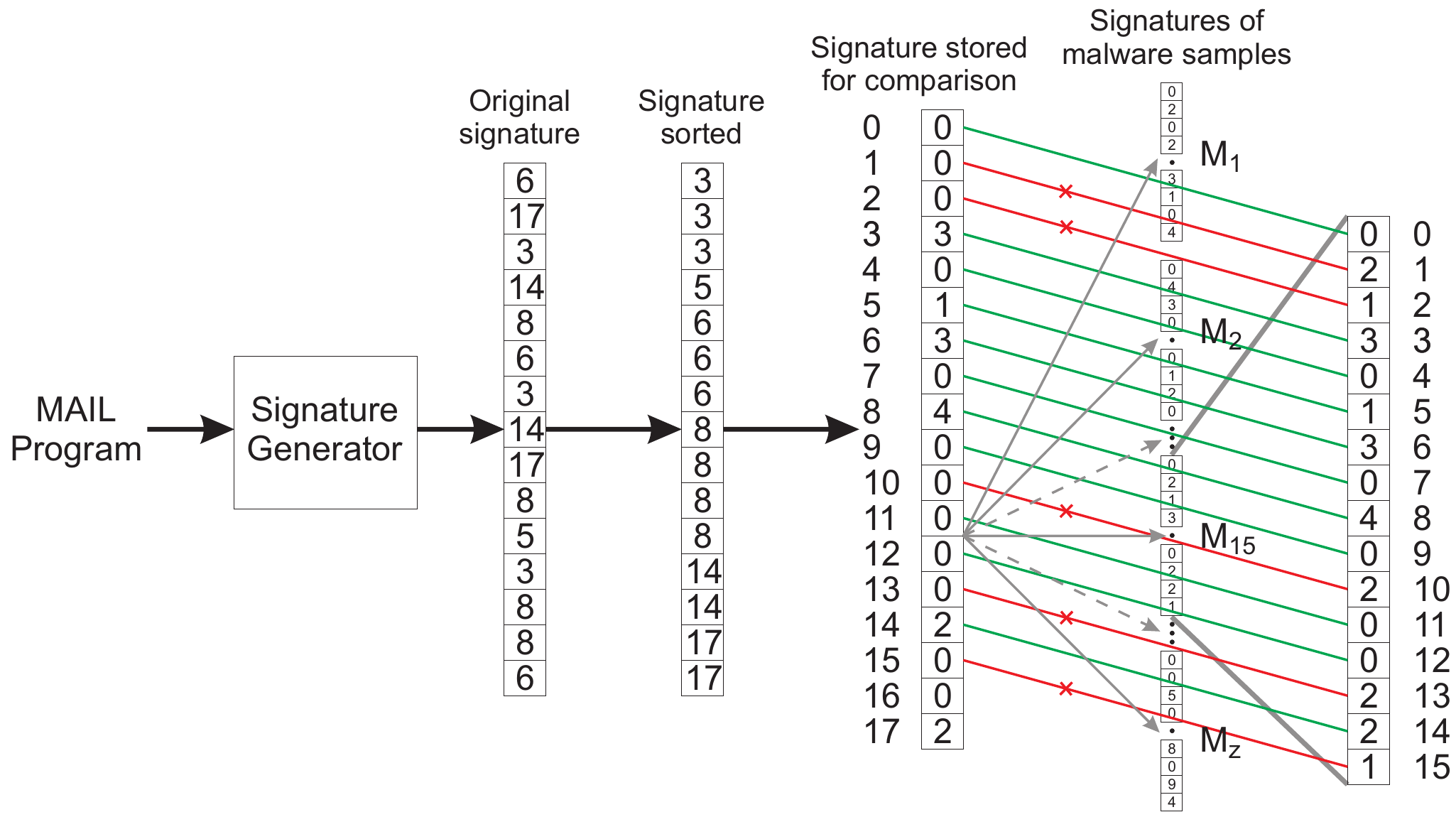}}
\caption{A successful SWOD signature matching.}
\label{fig:malware-detection-signatures-SWOD}
\end{figure}

After building the signature of a new program as explained above, it is compared with the signatures of all the training malware samples. In case of a match with any of the signatures we tag the new program as malware. 
An example of successful SWOD signature matching can be found in Fig.~\ref{fig:malware-detection-signatures-SWOD}, where 11 out of 16 (a threshold computed empirically) index-based array values of the program match with the malware program $M_{15}$.


\subsection{Malware Templates}\label{sec:mt}

The malware templates used for comparison are generated during a training phase that analyzes known malware and extracts their signatures as explained in Sections \ref{sec:sgt}, \ref{sec:acfg}, and \ref{sec:swod}. In this phase we find the threshold to be used for ACFG matching, and the differences between malware and benign samples' MAIL patterns to find the SWOD values as described in~\cite{SWOD-CFWeight-2014}. The threshold is the best similarity score of ACFGs matching that achieves the highest DR for the chosen dataset. The ACFG matching threshold computed (using 25\% of the dataset) for the dataset used in this paper is 70\%. The control flow weight of each MAIL pattern is computed, which is then used with the SWOD to compute the total weight of each MAIL pattern. 
The computed threshold and weight of each MAIL pattern are then used to build the \emph{Malware Templates} (malware signatures) shown in Figure~\ref{fig:DroidNative}.

\section{Evaluation}
In order to analyze the correctness and efficiency of {\dn} we performed a series of experiments of the system, comparing it to existing commercial systems as well as research systems.

\subsection{Experimental Setup}\label{sec:dataset}

\subsubsection{Dataset}

\begin{table}[tb]
\setlength{\tabcolsep}{5.3pt}
\centering
\caption{Class distribution of the 1240 malware samples, including whether they are native code or byte-code}
{
\scriptsize
\begin{tabular}{  l  c  c  } \hline
\textbf{Class/Family} & \textbf{Number of samples} & \textbf{Code Type} \\ \hline

CarrierIQ        & 8   & ARM \\ \hline
DroidPak         & 5   & Intel x86 \\ \hline
ChathookPtrace   & 2   & ARM \\ \hline
DroidKungFu1     & 35  & Bytecode \\ \hline
DroidKungFu2     & 30  & Bytecode \\ \hline
DroidKungFu3     & 310 & Bytecode \\ \hline
DroidKungFu4     & 95  & Bytecode \\ \hline
JSMSHider        & 15  & Bytecode \\ \hline
ADRD             & 21  & Bytecode \\ \hline
YZHC             & 22  & Bytecode \\ \hline
DroidDream       & 16  & Bytecode \\ \hline
GoldDream        & 45  & Bytecode \\ \hline
DroidDreamLight  & 46  & Bytecode \\ \hline
KMin             & 50  & Bytecode \\ \hline
Geinimi          & 68  & Bytecode \\ \hline
Pjapps           & 75  & Bytecode \\ \hline
BaseBridge       & 122 & Bytecode \\ \hline
AnserverBot      & 186 & Bytecode \\ \hline
Contagiominidump & 89  & Bytecode \\ \hline


\end{tabular}
}
\label{tab:data-set-classes}
\end{table}

Our dataset for the experiments consists of total 2240 Android applications. Of these, 1240 are Android malware programs collected from two different resources \cite{Dissecting-AM, Contagiominidump} and the other 1000 are benign programs containing Android 5.0 system programs, libraries and standard applications.

Table \ref{tab:data-set-classes} lists the class/family distribution of the 1240 malware samples. As can be seen, some of the malware in our set is native code based while others are bytecode based. \emph{CarrierIQ} is an adware that performs keystroke logging, location tracking and has the ability to intercept text messages. \emph{DroidPak} is an APK loader. It is a malware specifically designed to infect Android devices. \emph{ChathookPtrace} performs malicious code injection. The 4 \emph{DroidKungFu} variants make use of obfuscations to evade detection, such as different ways of storing the command and control server addresses and change of class names, etc. \emph{ADRD}, \emph{DroidDream}, \emph{DroidDreamLight} and \emph{BaseBridge} also contain number of variants through obfuscations. \emph{AnserverBot} employs anti-analysis techniques, such as, tampering of repackaged application; aggressively obfuscating its internal classes, methods and fields to make them humanly unreadable; and detection of some of the anti-malware programs. The samples collected from Contagiominidump~\cite{Contagiominidump} consists of differrent latest malware applications, such as \emph{RansomCollection}, \emph{FakeMart}, \emph{FakeJobOffer} and \emph{FakeAntiVirus}, etc.

In order to facilitate the possibility of other authors comparing their systems to {\dn}, we have made the source code (60,000+ lines of C/C++ code) of {\dn} and our testing dataset publicly available~\cite{DoubleBlind}.

\subsubsection{Test Platform}

All experiments were run on an Intel\textregistered{} Xeon\textregistered{} CPU E5-2670 @ 2.60GHz with 128 GB of RAM, running Ubuntu 14.04.1. Although such a powerful machine is not required for running DroidNative, it was used  in order to speed the results of n-fold testing, which requires a significant number of repetitions of experiments. The ART compiler, cross build on the above machine, was used to compile Android applications (malware and benign) to native code.


\subsection{N-Fold Cross Validation}
\label{sec:pm}

We use n-fold cross validation to estimate the performance of our technique. In n-fold cross validation the dataset is divided randomly into $n$ equal size subsets. $n - 1$ sets are used for training and the remaining set is used for testing. The cross validation process is then repeated $n$ times with each of the $n$ subsets used exactly once for validation. This results in very systematic, accurate testing results for a given dataset and removes the potential for skewed results based on the choice of which data is used for training and testing.\\
Before evaluating the proposed techniques, we first define the following evaluation metrics:

\begin{itemize}[leftmargin=*]
\item \textbf{DR}.  The Detection Rate, also called the true positive rate, corresponds to the percentage of samples correctly recognized as malware out of the total malware dataset.
\item \textbf{FPR}.  The False Positive Rate corresponds to  the percentage of samples incorrectly recognized as malware out of the total benign dataset.
\item \textbf{ROC}. The Relative Operating Characteristic curve is a graphical plot used to depict the performance of a binary classifier. It is a two dimensional plot, where DR is plotted on the Y-axis and FPR is plotted on the X-axis, and hence depicts the trade-offs between benefits (DR) and costs (FPR). We want a higher DR and a lower FPR, so a point in ROC space to the top left corner is desirable.
\item \textbf{AUC}.  The Area Under the ROC Curve is equal to the probability that a detector/classifier will detect a randomly chosen malware sample as malware and a randomly chosen benign sample as benign. 
For example, an AUC=1.0 means a perfect detector, i.e, the detector will always detect correctly, and an AUC=0.5 means the detector will detect correctly half of the time.
\end{itemize}

Through n-fold cross validation four experiments were conducted using two different subsets of the dataset. The first set consisted of 40 benign and 40 malware samples and the second set consisted of 350 benign and 330 malware samples. {\dn}-ACFG and {\dn}-SWOD were each tested with these two subsets of the dataset.

\begin{figure}[tb]
   \includegraphics[width=0.95\textwidth]{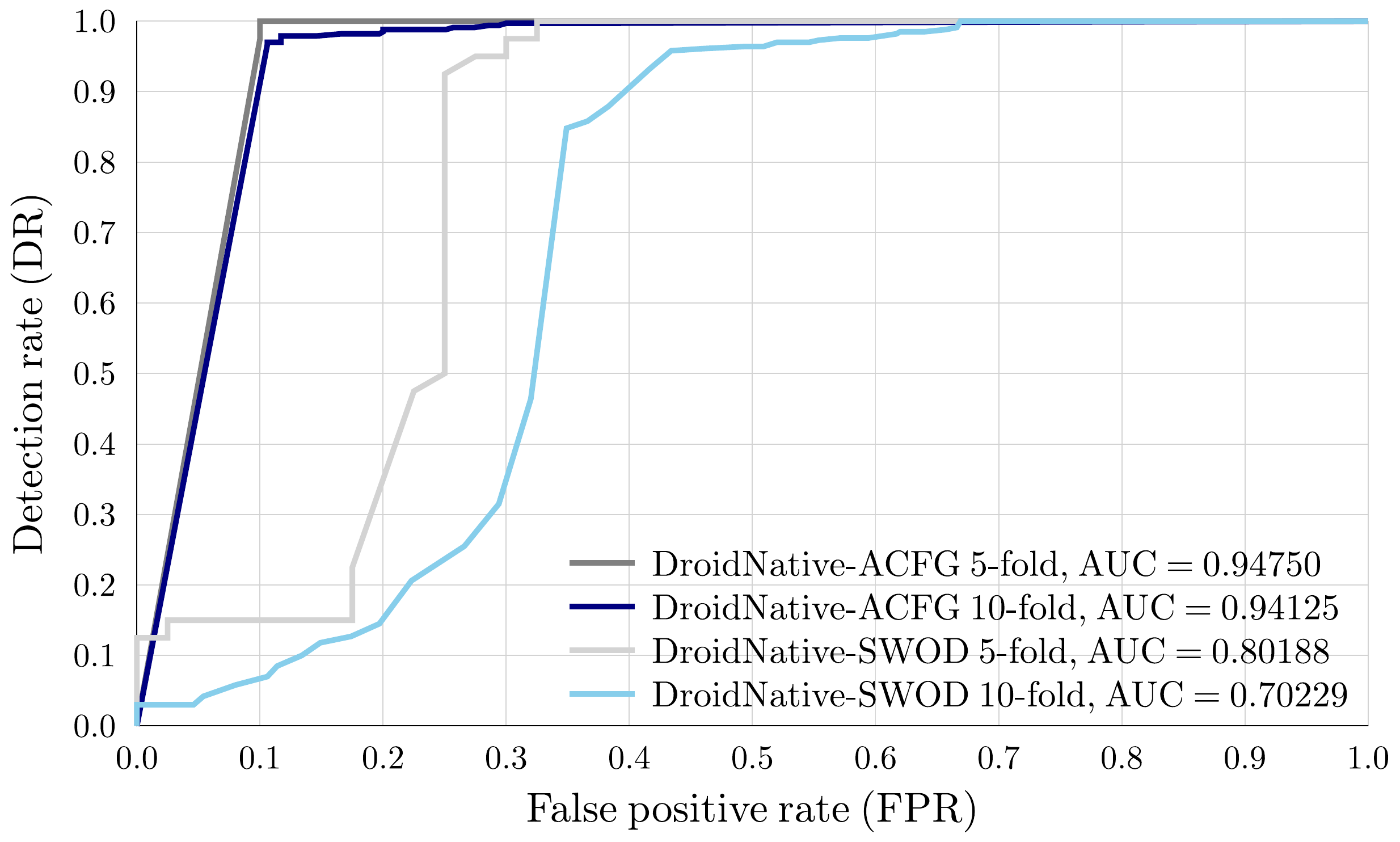}
	\caption{ROC graph plots of {\dn}'s ACFG and SWOD}
	\label{fig:ROC-plots-ACFG-SWOD}
\end{figure}

The ROC plot of both techniques using two different cross-validations can be seen in Fig.~\ref{fig:ROC-plots-ACFG-SWOD}. {\dn}-ACFG produces better results than {\dn}-SWOD. This difference is highlighted by the AUC of the two techniques. The AUC of {\dn}-ACFG range from 94.12 -- 94.75 while the AUC of {\dn}-SWOD range from 70.23 -- 80.19.

\subsection{Large-Scale Test}


To make our technique lightweight and general, we perform normalization of the binary code, as explained in Section \ref{sec:design}, that helps improve our detection results, i.e, the DR and the runtime. We could have used other binary analysis and optimization techniques such as method (API) calls, number of methods, loops and symbolic analysis etc. In general, these techniques require control and data flow analysis that are compute intensive and are not suitable for a real-time detector. Moreover, some of these analysis may not be accurate or possible in general to perform statically on machine code, as explained in Section~\ref{sec:design}.

To further reduce the runtime, the current implementation of {\dn} uses a simple binary decision tree classifier, which has proven exceptionally good for detecting malware variants. Currently the classifier used in {\dn} is only using commonalities among the malware samples. In the future we will be carrying out a study to also find commonalities among the benign samples, and between the malware and benign samples. These commonalities will be used to train the classifier that will help further reduce {\dn}'s FPR.


In order to demonstrate the scalability of our approach we also performed an experiment with a larger dataset of 2240 samples.

We trained {\dn} on 1000 malware samples and then performed detection on a set of 240 malware and 1000 benign samples.  The results of this test showed that {\dn} achieved a DR of 99.16\% and a FPR of 1.3\%, with an average detection time of 26.87 sec/sample. These results show that {\dn} is applicable to larger datasets.



\subsection{Comparison with Commercial Tools}
\label{sec:malware-variants-detection}

\begin{table*}[tb]
\setlength{\tabcolsep}{15pt}
\centering
\caption{Detection results of the 16 commercial anti-malware programs and {\dn} tested with 192 variants of 30 malware families.}
{
\scriptsize
\begin{tabular}{  l  c  c  c  c  c  c  c  c  } \hline

\textbf{Anti-malware} & \multicolumn{8}{  c  }{\textbf{Detection Rate (\%)}} \\
                      & NOP & CND & FND & RDI & RVD & RNF & RNM & Overall \\ \hline

\textbf{{\dn}}  &  100   & 100   & 100   & 100   & 96.43   & 100   & 100   & 99.48 \\ \hline
Kaspersky             &  92.85 & 92.85 & 92    & 92.59 & 96.43   & 93.1  & 92.59 & 93.22 \\ \hline
Sophos                &  92.85 & 92.85 & 92    & 85.18 & 96.43   & 93.1  & 92.59 & 92.18 \\ \hline
AVG                   &  78.57 & 85.71 & 92    & 92.59 & 78.57   & 93.1  & 92.59 & 87.5  \\ \hline
BitDefender           &  85.71 & 92.58 & 84    & 85.18 & 85.71   & 86.2  & 85.18 & 86.45 \\ \hline
DrWeb                 &  85.71 & 53.57 & 84    & 85.18 & 89.28   & 79.31 & 85.18 & 80.31 \\ \hline
ESET                  &  89.28 & 53.57 & 88    & 88.88 & 17.85   & 89.65 & 85.18 & 72.91 \\ \hline
Microsoft             &  39.28 & 75    & 32    & 33.33 & 32.14   & 34.48 & 33.33 & 40.1  \\ \hline
VIPRE                 &  17.85 & 57.14 & 16    & 14.81 & 14.28   & 13.79 & 14.81 & 21.35 \\ \hline
Symantec              &  14.28 & 53.57 & 8     & 3.7   & 7.14    & 10.34 & 7.4   & 15.1  \\ \hline
Qihoo-360             &  10.71 & 53.57 & 8     & 7.4   & 7.14    & 6.89  & 7.4   & 14.58 \\ \hline
Fortinet              &  10.71 & 53.57 & 8     & 7.4   & 7.14    & 6.89  & 7.4   & 14.58 \\ \hline
TrendMicro            &  3.57  & 53.57 & 4     & 7.4   & 3.57    & 3.57  & 3.7   & 11.45 \\ \hline
McAfee                &  7.14  & 53.57 & 4     & 3.7   & 3.57    & 3.57  & 3.7   & 11.45 \\ \hline
TotalDefense          &  7.14  & 53.57 & 0     & 0     & 0       & 0     & 0     & 8.85  \\ \hline
Malwarebytes          &  3.57  & 53.57 & 0     & 0     & 0       & 0     & 0     & 8.33  \\ \hline
Panda                 &  3.57  & 53.57 & 0     & 0     & 0       & 0     & 0     & 8.33  \\ \hline

\end{tabular}
}
\label{tab:comparison-commercial-malware-variants}
\end{table*}

In order to compare {\dn} to a variety of commercial tools we generated variants of known malware in order to test their effectiveness. We selected 30 different families (class/type) of Android malware from~\cite{Contagiominidump}, and used DroidChameleon~\cite{DroidChameleon} to generate 210 malware variants based on them. DroidChameleon applies the following 7 obfuscations: $\mathcal{NOP}$ = no-operation insertion; $\mathcal{CND}$ = call indirection; $\mathcal{FND}$ = function indirection; $\mathcal{RDI}$ = removing debug information; $\mathcal{RVD}$ = reversing order; $\mathcal{RNF}$ = renaming fields; $\mathcal{RNM}$ = renaming methods. 

In total, we generated 210 variants, but 18 of them were not able to run on the new ART runtime.  DroidChameleon was implemented and tested using the Dalvik-VM, and apparently some of the variants result in bytecode that is consider invalid by ART.  As such, we excluded these variants, leaving 192 samples.

We tested 16 commercial anti-malware programs and {\dn} with these 192 Android malware variants. For testing {\dn} we included only the original 30 malware samples as the training dataset and the other 192 variants and 20 benign samples as the testing dataset. {\dn} was able to properly classify all benign samples and 191 out of 192 malware samples. 
This is an overall DR = 99.48\% and FPR = 0\%. The DR of {\dn} reported here is the combination of both {\dn}-ACFG and {\dn}-SWOD, i.e, the malware samples detected by both.

Overall, only one malware variant sample was not detected by {\dn}, and it belonged to class $\mathcal{RVD}$. This class/obfuscation inserts \emph{goto} instructions that changes the CFG of a method/program. After further analysis of the original and the variant malware samples we notice that the difference (the number of blocks and the control flow change per ACFG) in ACFGs of the two samples was more than 50\%. As we discuss later in Section~\ref{sec:limitations}, {\dn} may not be able to detect a malware employing excessive control flow obfuscations. Therefore, {\dn} was not able to detect this variant as a copy of the original malware sample.

We then tested a variety of commercial anti-malware systems with the same 192 variants.  The results can be found in Table~\ref{tab:comparison-commercial-malware-variants}. The majority (11/16) of the commercial anti-malware tools performed very poorly, with overall DRs ranging from 8.33\% -- 72.91\%.  Five of the tools performed significantly better, but were still not able to match the results of {\dn}. Overall, the commercial tools did better in our test than in previous reports~\cite{DroidChameleon,ADAM,Obfuscation-Faruki}, which shows that their detection techniques are continuously being improved.



\subsection{Comparison with Existing Research}\label{sec:comparison}

\begin{table*}[tb]
\setlength{\tabcolsep}{10pt}
\centering
\caption{Comparison with other malware detection techniques discussed in Section \ref{sec:related_works}. Out of the 1240 malware samples, 1000 were used for training and 240 for testing.}
{
\small
\begin{tabular}{  l  c  c  c  c  c  } \hline

\textbf{Technique} & \textbf{DR} & \textbf{FPR} & \textbf{Dataset size} & \textbf{Runtime} & \textbf{Android} \\[-0.7ex]
\textbf{}          & \textbf{}   & \textbf{}    & \textbf{benign/}      & \textbf{per}     & \textbf{native}    \\[-0.7ex]
\textbf{}          & \textbf{}   & \textbf{}    & \textbf{malware}      & \textbf{sample}  & \textbf{binary}    \\ \hline

\textbf{{\dn}}                    & 99.16\% & 1.3\%   & 1000 / 1240 & 26.87 sec & \ding{51} \\ \hline
DroidSift   \cite{DroidSIFT-AM}   & 93\%    & 5.15\%  & 2100 / 193  & 175.8 sec & \ding{55} \\ \hline
DroidLegacy \cite{DroidLegacy-AM} & 92.73\% & 20.83\% & 48 / 743    & \ding{55} & \ding{55} \\ \hline
AndroSimilar \cite{AndroSimilar}  & 76.48\% & 1.46\%  & 21132 / 455 & \ding{55} & \ding{55} \\ \hline

\end{tabular}
}
\label{tab:comparison}
\end{table*}

In this Section we compare three \cite{DroidSIFT-AM,DroidLegacy-AM,AndroSimilar} state of the art research efforts carried out in 2014 that use machine learning for Android malware variant detection. Very few works report the runtime of their detector, and to compare the runtime of our detector we selected DroidSift~\cite{DroidSIFT-AM}. We selected DroidLegacy~\cite{DroidLegacy-AM}, because of their use of n-fold cross validation for evaluating the performance of their detector. Like {\dn}, DroidSift and AndroSimilar~\cite{AndroSimilar} also test their techniques on malware variants generated through obfuscations.

Table \ref{tab:comparison} gives a comparison of {\dn} with these three approaches. For DroidSift, we have only reported the DR of their malware variant detector, because their anomaly detector does not provide automation of malware detection. In DroidLegacy the same 48 applications were used in each of the 10-folds (confirmed through e-mail by the authors). During the n-fold testing performed in the paper there were a total of 10 different benign applications that were detected as malware, therefore, according to our computation the actual FPR = 10/48 = 0.2083.

The application of DroidSift and AndroSimilar to general malware and impartiality towards the testing dataset is not validated through n-fold cross validation or other such testing approaches. Their testing may also suffer from the possibility of data overfitting. DroidLegacy needs a larger number of benign samples (more than 48) than tested in the paper for further validation. None of the other techniques can analyze Android native binaries. Out of the three techniques, only AndroSimilar has the potential to be used in a real-time detector but has poor DR results. {\dn}, on the other hand, is automated, fast, platform independent, and can analyze and detect Android native code malware.

The techniques proposed in DroidSift and DroidLegacy are very similar to {\dn}-ACFG. The difference is that {\dn}-ACFG captures control flow with patterns of the Android application's native binary code, whereas DroidSift and DroidLegacy capture the API call graph from the Android application's bytecode, and may not be able to detect malware applications that obfuscate by using fake API calls, hiding API calls (e.g, using class loading to load an API at runtime), inlining APIs (e.g, instead of calling an external API, include the complete API code inside the application), and reflection  (i.e, creation of programmatic class instances and/or method invocation using the literal strings, and a subsequent encryption of the class/method name can make it impossible for any static analysis to recover the call).

In DroidSift, during training the malware graphs that were common to the benign graphs are removed, i.e, there can be malware applications whose API calls are similar to the benign applications; hence their technique may not detect such malware applications. The anomaly detector proposed in DroidSift is based on the behavior of known benign applications. Keeping track of benign applications is impractical in real life, and also makes it more complex in terms of resources (time and memory) required than the technique proposed in this paper.
Hence, DroidSift is much less efficient than {\dn} from a performance perspective. The concept used in DroidSift for anomaly detection is similar to the use of \emph{Whitelists}, and will suffer from the same issues, such as difficultly to collect and delay between discovery and updating. Our technique is based on the behavior of known malware applications, hence is more practical and lighter in terms of runtime. Unlike {\dn}, however, their anomaly detector may be able to detect zero-day malware.

{\dn}-SWOD is similar to AndroSimilar \cite{AndroSimilar}, which is based on SDHash \cite{SDHash} and is most likely to detect very similar objects, and that's why it produces low DRs. {\dn}-SWOD is a trade off between accuracy and efficiency, and detects similar objects at a coarse level, therefore it produces high DRs.

In terms of variant detection, DroidSift was tested with 23 variants of DroidDream, only 1 family (listed in Table \ref{tab:data-set-classes}) of malware, and achieved an overall DR = 100\%. They used one of the obfuscations described in \cite{DroidChameleon} to generate these variants, but it's not clear what specific obfuscation was used. AndroSimilar was tested with 1013 variants of 33 families (most of them are listed in Table \ref{tab:data-set-classes}) of malware, and achieved an overall DR = 80.65\%. They used the following 4 obfuscations, \emph{method renaming}, \emph{junk method insertion}, \emph{goto insertion} and \emph{string encryption}, to generate these variants. {\dn} is tested with 192 variants of 30 families of malware generated using 7 types of obfuscation, and achieved an overall DR = 99.84\%.

In terms of performance, only DroidSift reported runtime results. Although we acknowledge that it is less than ideal to directly compare performance numbers of two different systems tested on completely different hardware, the fact that {\dn} is {\raise.17ex\hbox{$\scriptstyle\sim$}}6.5 times faster is a stronger indicator of the system's significant performance advantage. More thorough performance testing would require access to the source code of the other systems, which was not available to us.

\subsection{Limitations}\label{sec:limitations}

{\dn} has some limitations, discussed in this section.

Like all static analysis techniques, {\dn} requires that the application's malicious code be available for static analysis.  Techniques such as packing (compressing and encrypting the code) and applications that download their 
malicious code upon initial execution cannot be properly analyzed by the system.

{\dn} may not be able to detect true zero-day malware. {\dn} excels at detecting variants of malware that has been previously seen, and will only detect a zero-day malware if its control structure is similar (upto a threshold) to an existing malware sample in the saved/training database.

{\dn} may not be able to detect a malware employing excessive control flow obfuscations. Excessive here means changing the control flow of a program beyond a certain percentage (threshold). In general, such threshold is difficult to find, but is calculated for each different dataset. These control flow obfuscations can be: 
\emph{control flow flattening}, obscuring the control flow logic of a program by flattening the CFG; 
\emph{irreducible flow graphs}, reducing these flow graphs looses the original CFG; 
\emph{method inlining}, replacing a method call by the entire method body, which may increase the number of CFGs in the program; 
\emph{branch function/code}, obscuring the CFG by replacing the branches by the address of a function/code; 
and \emph{jump tables}, artificial jump table or jumps in the table are created to change the CFG.

The pattern matching of two ACFGs may fail if the malware variant obfuscates a statement in a basic block in a way that changes its MAIL pattern. {\dn} performs exact (100\%) pattern matching for two ACFGs to match. To improve {\dn}'s resilient to such obfuscations, in the future we will use a threshold for pattern matching. We will also investigate other pattern matching techniques, such as a statement dependency graph or assigning one pattern to multiple statements of different type etc, to improve this resiliency.




\section{Related Work}\label{sec:related_works}

\subsection{Native Code Analysis for Other Platforms}

There are several efforts for detecting malware native binaries for standard PCs. We briefly discuss some of these techniques. We also discuss the use of intermediate languages for detecting malware native binaries.

In \cite{CFA-Model-Checking-MD-1}, a method is presented that uses model-checking to detect malware variants. They build a CFG of a binary program, which contains information about the register and the memory location values at each control point of the program. Model-checking is time consuming and can run out of memory. Times reported in the paper range from few seconds (10 instructions) to over 250 seconds (10000 instructions). \cite{CFA-1} presents a method that uses CFGs for visualizing the control structure and representing the semantic aspects of a program. They extend the CFG with extracted API calls to have more information about the executable program. \cite{Code-Graph-MD} proposes a technique that checks similarities of call graphs (semantic signatures) to detect malware variants. Only system calls are  extracted from the binary to build the call graph. The call graph is reduced to reduce the processing time, but this also reduced the accuracy of the detector. \cite{Value-Set-MD-1} uses value set analysis (VSA) to detect malware variants. Value set analysis is a static analysis technique that keeps track of the propagation and changes of values throughout an executable. VSA is an expensive technique and cannot be used for real-time malware detection.

These techniques are compute intensive, and attempts to reduce the processing time tend to produce poor DRs, cannot handle smaller size malware, and are not suitable for real-time detection. Other techniques \cite{Opcode-Graph-MD,Opcode-Histogram,Opcode-HMM-MD,Opcode-SD-MD} that mostly use opcode-based analysis for detecting malware variants have the potential to be used for real-time malware detection, but have several issues:
The frequencies of opcodes can change by using different compilers, compiler optimizations, architectures and operating systems; obfuscations introduced can change the opcode distributions; selecting too many features (patterns) for detection results in a high detection rate but also increases the runtime.

\subsubsection{Intermediate Languages}

REIL~\cite{reil2009} is being used for manual malware analysis and detection. Unhandled native instructions are replaced with NOP instructions which may introduce inaccuracies in disassembling. Furthermore, REIL does not translate FPU, MMX and SSE instructions, nor any privileged instructions, because these instructions according to the authors are not yet being used to exploit security vulnerabilities. SAIL~\cite{SAIL} represents a CFG of the program under analysis. A node in the CFG contains only a single SAIL instruction, which can make the number of nodes in the CFG extremely large and therefore can make analysis excessively slow for larger binary programs. In VINE~\cite{bitblaze}, the final translated instructions have all the side effects explicitly exposed as VINE instructions, that makes this approach general but also difficult to maintain platform independence and less efficient for specific security applications such as malware detection. CFGO-IL~\cite{CFGO-IL} simplifies transformation of a program in the x86 assembly language to a CFG. By exposing all the side effects in an instruction, CFGO-IL faces the same problem as VINE-IL. Furthermore, the size of a CFGO-IL program tends to be much larger than the original assembly program. WIRE \cite{wire} is a new intermediate language and like MAIL is specifically designed for malware analysis. WIRE does not explicitly specify indirect jumps, making malware detection more complicated. Furthermore, the authors do not mention anything about the side effects of the WIRE instructions, and is not clear how the language is used to automate the malware analysis and detection process.

In contrast to other languages: side-effects are avoided in MAIL, making the language much simpler and providing the basis for efficient malware detection; control flow with patterns in MAIL provide further optimization and automation of malware detection; publicly available formal model and tool makes MAIL easier to use.

\subsection{Detection of Android Malware Variants}
Because of the popularity of Android devices, there are several research efforts on malware detection for Android. For a comprehensive survey of Android malware detection techniques the reader is referred to \cite{Survey-Android-Security}. Here, we only discuss the three \cite{DroidSIFT-AM,DroidLegacy-AM,AndroSimilar} research efforts that are also used for comparison in Section~\ref{sec:comparison}.

\cite{DroidSIFT-AM} detects unknown malware using anomaly detection. They generate an API call graph and assign weights based on the data dependency and frequency of security related API calls. A graph of an application is matched against a graph database of benign applications, and if a matching graph is not found then an anomaly is reported.
These reports are then sent to a human expert for verification.
For known malware, a signature is generated for each malware sample using the similarity score of the graphs. A classifier is used for signature detection, and is called the malware variant detector.


In general, changing (obfuscating) control flow patterns is more difficult (i.e, it needs a comprehensive change in the program) than changing just API call patterns of a program to evade detection. ACG based techniques look for specific API call patterns (including call sequences) in Android malware programs for there detection, which may also be present in Android benign applications that are protected against reverse engineering attacks. These API call patterns can be packing/unpacking, calling a remote server for encryption/decryption, dynamic loading, and system calls, etc. Moreover, sometimes, e.g, for stripped native binaries of Android, it becomes impossible to build a correct ACG of a program.

\cite{DroidLegacy-AM} uses API call graph signatures and machine learning to identify piggybacked applications. Piggybacking can be used to inject malicious code into a benign application. First a signature is generated of a benign application and then it is used to identify if another application is a piggyback of this application or not. The technique used has the same limitations as discussed above about the detection schemes based on API calls.

\cite{AndroSimilar} is based on SDHash \cite{SDHash}, a statistical approach for selecting fingerprinting features. SDHash relies on entropy estimates and an empirical study, and is most likely to pick features unique to a data object. Therefore the results in the paper have a better false positive rate. Although the FPR reported in \cite{AndroSimilar} is low (1.46\%), because of the SDHash technique used, whose main goal is to detect very similar data objects, the ability of detecting malware variants is much lower than the technique proposed in this paper.

\subsection{App Analysis on iOS}
One other effort \cite{Privacy-IOS} targets Apple's iOS, and performs static binary analysis by building a CFG from the Objective-C binaries. The main goal of the work is to detect privacy leaks and not malware analysis and detection. They perform complex data flow analysis such as backward slicing and others which are time consuming and are not suitable for real-time malware detection. There are some differences between Android applications and Apple iOS applications, such as the use of C, C++ and Java in Android and Objective-C in iOS, and the use of components in Android and use of message passing in iOS, etc. Because of these differences, binary analysis of these applications have significant differences and challenges.

\section{Conclusion}\label{sec:conclusion}

In this paper we have proposed \emph{{\dn}} for the detection of both bytecode and native code Android malware variants. {\dn} uses control flow with patterns, implements and adapts the two techniques ACFG and SWOD to reduce the effect of obfuscations; uses the language MAIL to provide automation and platform independence; and has the potential to be used for real-time malware detection. To the best of our knowledge, none of the existing static analysis techniques deal with the detection of Android native code malware, and this is the first research effort to detect such malware. Through n-fold cross validation and other test evaluations, {\dn} shows superior results for the detection of Android native code and malware variants compared to the other research efforts and the commercial tools.




\end{document}